# Germanium and Silicon Detectors for WIMP Searches: Key Experimental Insights

M. Mirzakhani, [a]* S. Maludze [a]

[a]*Department of Physics and Astronomy, Texas A&M University, College Station, TX, 77843*

[*]*Mahdi_mirzakhani1995@tamu.edu*

## Abstract

Dark matter plays a crucial role in our comprehension of the universe, but its mysterious nature poses challenges for direct detection. A primary obstacle in detecting dark matter is distinguishing genuine signals from the prevailing electromagnetic background. Germanium and Silicon detectors have emerged as effective instruments in the pursuit of dark matter detection. Their minimal radioactive backgrounds, substantial active volumes, and efficient rejection mechanisms have significantly advanced our understanding of dark matter and its scattering cross section limits. Numerous experiments employing these detectors have yielded valuable insight into the properties of dark matter. Scientists have investigated potential dark matter candidates like Weakly Interacting Massive Particles (WIMPs) and evaluated their probability of being dark matter particles based on observed scattering cross sections. This review consolidates the findings from significant experiments, encompassing possible candidates and their likelihood of being dark matter particles. It also recognizes the limitations of dark matter scattering cross section by assessing progress in this field, detector technologies, experimental outcomes, and future prospects.

*Keywords:* Dark matter, Germanium detector, Silicon Detector, Nuclear recoil energy, Nucleon-scattering cross-section, Low temperature

## I. Introduction

It is believed that dark matter constitutes about 85% of the universe's matter content [1]. Its presence is inferred from its gravitational influence on normal matter, radiation, and the cosmic structure, yet it eludes direct observation because it does not interact with electromagnetic and strong forces [2]. Among the various potential forms of dark matter, Weakly Interacting Massive Particles (WIMPs) stand out as prime candidates [3]. To detect WIMPs, experiments have frequently employed germanium and silicon detectors, which are favored for their low levels of background noise, high purity, and capability to function at extremely low temperatures [4].

The following sections provide an overview of dark matter candidates and how they interact. It also explores the detector technologies, and their readout tools used in searching for rare events with germanium and silicon detectors. Furthermore, it highlights significant results from various collaborations and experiments. Lastly, the section briefly discusses the challenges and prospects in this field.

## II. Theoretical Background

### *II.A. Dark Matter Candidates*

**WIMP**s: Massive particles that interact weakly via gravity and do not participate in electromagnetic or strong interaction, known as WIMPs [5], are hypothesized to constitute the bulk of the universe's matter [6,7]. The evidence supporting the existence of WIMPs is becoming increasingly compelling, with the latest support coming from a combination of the Wilkinson Microwave Anisotropy Probe findings with research on large-scale structure formation, as observed by the Sloan Digital Sky Survey, and data from supernova redshift surveys [8–12]. An acceptable theory for how WIMPs are spread throughout our galaxy, which aligns with the observed rotational velocities of spiral galaxies [13], proposes that they form an approximately isothermal spherical halo encircling the galaxy, averaging velocities around 230 kilometers per second [2,14]. Specifically, dark matter with a mass less than a $\mathrm{GeV/c^2}$, which interacts with Standard Model particles via a novel force carrier, presents a credible option to the traditional WIMP theory [15–17]. Bosonic dark matter, such as dark photons, with masses on the scale of $\mathrm{eV/c^2}$, as well as fermionic dark matter with masses in $\mathrm{MeV/c^2}$ range, which would be the lightest particle in a novel force field [18–22], are both able to account for the observed abundance of dark matter in the universe while remaining undetected by existing searches [23–25].

**Axion**s: Particles which are a particularly compelling category of dark matter candidates due to their strong theoretical foundation and the prospect of significant experimental breakthroughs, possibly leading to their discovery within the coming ten years. These lightweight pseudoscalar particles are predicted by numerous theoretical extensions of the Standard Model [26]. A prominent instance of an axion is the QCD (Quantum ChromoDynamic) axion, which emerges from the Peccei-Quinn mechanism that addresses the strong-CP (**C**harge conjugation symmetry and **P**arity symmetry) problem [27–30].

**Sterile Neutrinos:** These are posited as elusive particles that do not interact via the electromagnetic or strong forces, being neutral under the Standard Model's charges. They may, however, have the capacity to interact with regular matter by mixing with the active neutrinos that participate in weak force interactions [31]. Notably, heavier sterile neutrinos might play a significant role in cosmology, such as being candidates for dark matter at the keV scale or contributing to Leptogenesis, a process that might account for the universe's matter-antimatter imbalance [32]. Intriguingly, several experimental discrepancies, like those observed in the Liquid Scintillator Neutrino Detector (LSND) and MiniBooNE experiments, as well as reactor and gallium anomalies, hint at sterile neutrinos' existence since they challenge the predictions made by the Standard Model's three-neutrino framework [33].

### *II.B. Interaction Mechanisms*

**Elastic Scattering**: The domain of dark matter direct detection is focused on a broad spectrum of experiments, primarily aimed at the discovery of WIMPs. As explained before, these particles are hypothesized to explain the relic dark matter density observed in the universe, believed to have formed in the early universe in thermal equilibrium with other particles [5,33–47]. The detection strategy in these experiments revolves around the elastic scattering of WIMPs off atomic nuclei, with the observation of nucleus recoil as the primary signal [49]. Theoretical models that support these rates are being tested by experiments that are now sensitive enough to detect them. These models often rely on new physics beyond the Standard Model, like supersymmetry or theories involving extra dimensions, which are linked to electroweak symmetry breaking at energy scales considered low for particle physics, such as 100 GeV. The energy transfer involved in these direct detection events is very small, often just a few hundred MeV [50].

**Inelastic Scattering**: While elastic scattering is the common search strategy, inelastic scattering also plays a role [42,43,51,52]. In such cases, the WIMP interaction leaves the nucleus in an excited state, which then emits a gamma-ray upon de-excitation, providing an additional signal [53]. For example, the inelastic scattering of a WIMP with a $^{129}$Xe nucleus leads to energy release primarily through electronic recoil as the emitted photon dissipates energy within the liquid xenon of the detector, after the nuclear recoil event [54].

## III. Germanium and Silicon Detectors

### *III.A. Detector Technologies*

**Cryogenic Detectors**: In the past decade, the pursuit of dark matter has spurred the advancement of cryogenic detectors by various research groups [55–59], inspired by the compelling advantage that at extremely low temperatures, these detectors exhibit a heat capacity that adheres to the $T^3$ at Debye law, allowing for highly sensitive calorimetric measurements of minuscule energy deposits. Impressively, these detectors have achieved the capability to detect recoil energies below 1 keV. For instance, a 1 keV energy deposit in a 100 g cryogenic detector at 10 mK can cause a measurable temperature to rise of about 1 μK, detectable with standard electronics. Moreover, the energy needed to excite an elementary phonon in these detectors is less than 1 meV, a fraction of the energy required in traditional detectors like semiconductors or scintillators, paving the way for unparalleled sensitivity and energy resolution in dark matter detection [57–60]. The fundamental resolution of these detectors can be approximated by the thermodynamic fluctuations in the energy of the detector:

$$\Delta E_{FWHM} \approx 2.35 \sqrt{k_B C T^2}, \quad (1)$$

Where $k_B$ is the Boltzmann constant, C is the heat capacity of the detector, and T is the temperature. This resolution lies in the tens of electron-volt range for a 100 g detector at a temperature of 20 mK [61].

**Ionization Detectors**: Charge-phonon detectors utilize a discrimination technique that differentiates between electron recoils and nuclear recoils by measuring their distinct ionization efficiencies (Figure 1), known as the quenching factor. This factor is determined through experimental procedures involving neutron sources and neutron beams with specific tagging. Theoretical predictions for the quenching factor's energy dependency are made using phenomenological models like the Lindhard model [62]. For instance, in germanium detectors operating at low temperatures, it takes roughly 2.9 eV of electron energy to create an electron-hole pair due to the material's gap energy of about 0.7 eV [61], So the ionization efficiency for nuclear recoils is usually about three times less than that for electron recoils and varies depending on the energy deposited. By employing cooled Field-Effect Transistors (FETs) and SQUID (Superconducting Quantum Interference Device) electronics, these detectors achieve effective discrimination capabilities for recoil energies down to approximately 10 keV [61]. Building on the early successes with small silicon detectors that showcased the potential of charge-phonon discrimination [63], two significant dark matter detection experiments, CDMS [64] and EDELWEISS [65], have since incorporated this technology.

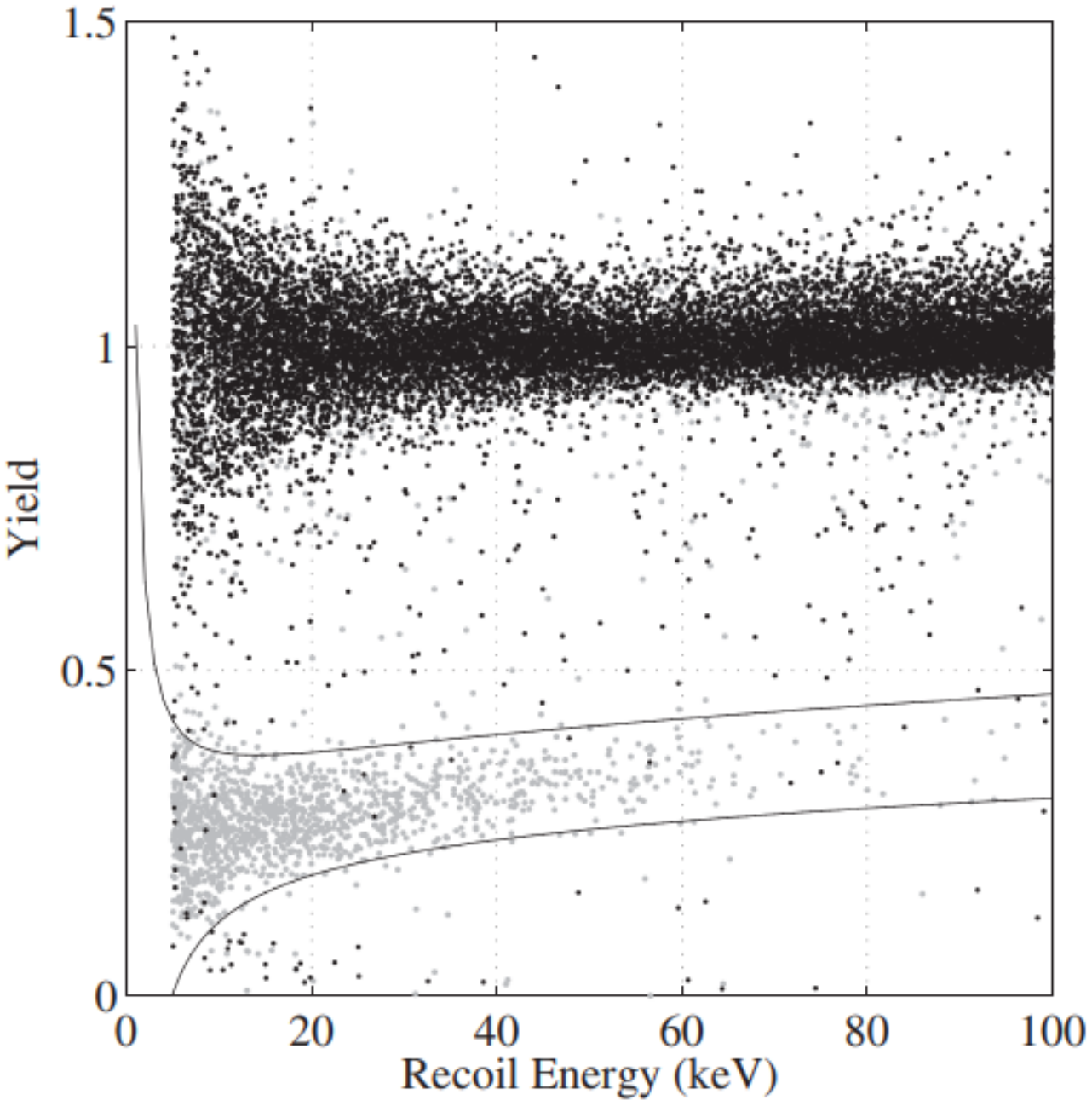

FIG. 1. Ionization yield as a function of recoil energy for a germanium detector (detailed in reference [66]). Calibration incidents involving only gamma emissions from a $^{133}Ba$ source is represented by black points, while calibration events involving both gamma and neutron emissions from a $^{252}Cf$ source are denoted by gray points. The higher cluster of events constitutes the bulk electron recoils, delineating what is known as the "electron-recoil band." Conversely, the lower cluster of events consists of nuclear recoils, which establish the "nuclear-recoil band."

**Scintillation Detectors**: These detectors operate by detecting light produced when a dark matter particle, with mass around the $\mathrm{MeV/c^2}$ scale, interacts with an electron in a crystal, transferring it from the valence to the conduction band, thus generating an electron-hole pair as is shown in Figure 2 [17]. Typically, the dark matter particle imparts a few eV to the electron, although it can occasionally transfer more energy at a lower probability [25]. When the electron has enough recoil energy, it can induce the creation of more electron-hole pairs as it settles to the bottom of the conduction band. In the scintillator, these pairs can recombine or be trapped by impurities that were added intentionally, leading to the emission of photons. The large cross-section for nuclear recoils is generally another hypothesis for interaction of dark matter with scintillator crystals through nuclear scattering. This larger cross-section means that nuclear recoils are more likely to occur and hence provide a more probable detection route. If the scintillator efficiently converts these interactions into light (high radiative efficiency), the number of emitted photons will reflect the number of electron-hole pairs created. These photons are then can be captured and converted into an electrical signal by photodetectors placed around the scintillating material [67].

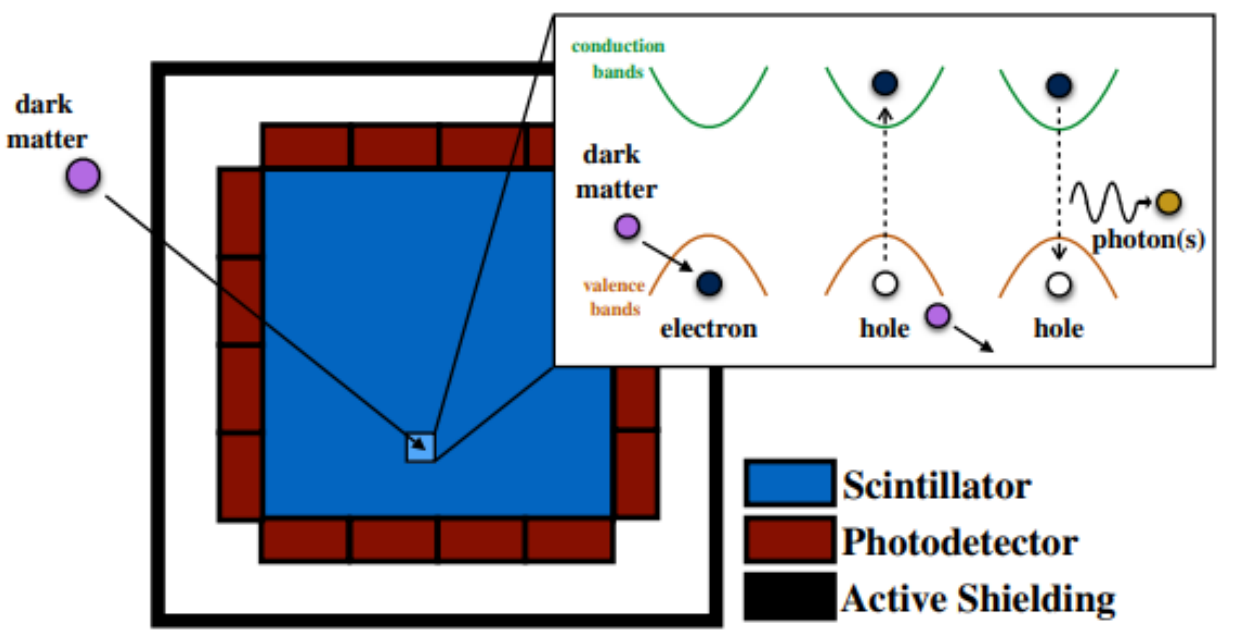

FIG. 2. When a dark matter (DM) particle collides with an electron in a scintillating crystal, it excites the electron to a higher energy state. As the electron returns to its ground state, it emits scintillation photons, which are then detected by an array of photodetectors surrounding the crystal. This setup here is protected by an active shield to block out environmental noise. Unlike methods that directly detect charge in semiconductor targets, this scintillation approach does not require an electric field, which helps to minimize or remove various detector-specific background interferences [67]. "Reproduced with permission from S. Derenzo, R. Essig, A. Massari, A. Soto, and T. T. Yu, Direct Detection of Sub-GeV Dark Matter with Scintillating Targets, Phys. Rev. D 96, 016026 (2017). Copyright 1998, American Physical Society."

### *III.B. Readout Technologies*

**Phonon Sensors:** They can detect non-thermal phonons generated by particle interactions in absorptive material. These phonons must first decay to a lower energy state and thermalize before the resulting temperature change ($\Delta T$) can be measured [68]. The thermalization process and the lengthy pulse recovery time limit the operational rate of these thermal calorimeters to only a few events per second. The most prevalent types of phonon sensors are resistive thermometers, such as semiconductor thermistors [69–71] and superconducting transition edge sensors (TES), which detect changes in temperature through corresponding changes in resistance [72,73]. However, these thermometers are subject to inherent Johnson noise and are dissipative; they require power to operate, which leads to additional heating of the calorimeter through Joule heating.

**Quasiparticle Detection:** Superconducting detectors operate on the principle of Cooper pair disruption and the generation of quasi-particles. These quasi-particles, produced when X-rays are absorbed or as a result of a passing particle's energy loss in a superconducting material, can be detected using a Superconducting Tunnel Junction (STJ), which functions similarly to the better-known Josephson junction [74–76]. By applying the right voltage across an STJ, the tunneling current reflects the number of quasi-particles created, which exceeds the thermal background at very low temperatures, less than 0.1 times the critical temperature ($T_c$). STJ arrays are also employed to detect energetic, non-thermal phonons in dielectric or superconducting absorbers.

Advancements in detection technology have led to Microwave Kinetic Inductance Detector (MKID), which uses a frequency-domain method for easier multiplexing and simplifies both the detector array and its readout systems. Additionally, another method involves using small Superheated Superconducting Granules (SSG) within an external magnetic field for detection purposes [77].

### *3.3. Operating Environment*

**Shielding**: In particle physics experiments are tailored for each setup but generally aim to reduce false signals and background noise. This includes protection against cosmic ray-induced neutrons, muons from cosmic rays interacting with nearby rocks or the shield itself, as well as gamma rays and neutrons emitted by radioactive elements in the surrounding environment. For example, Conducting experiments deep underground, equivalent to 2090 meters of water, can diminish the influx of surface muons by a factor of 50,000 [78,79]. An example of shielding in CDMS II is presented in Figure 3.

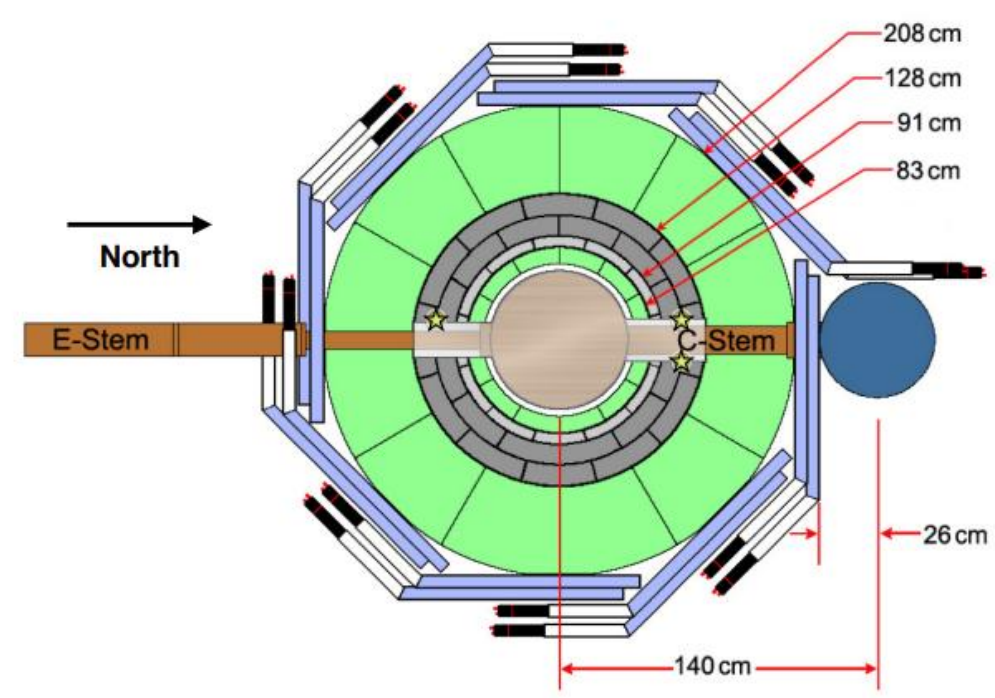

FIG. 3. The CDMS II device is designed with multiple layers of shielding, viewed from the top, with specified points for calibration sources (★). It features an external layer of staggered muon veto panels for detecting unwanted muon particles, encased in a green polyethylene layer to reduce neutron interference. This is followed by a protective layer of low-radioactivity gray lead bricks and a further layer of ancient lead with even lower radioactivity. Another layer of green polyethylene lies within, and the core is safeguarded by a mu-metal shield that prevents magnetic interference, wrapping around the bronze-colored copper cryostat cans. The mu-metal barrier also includes pathways that allow for electronic readouts and connections to the cooling system of the apparatus [80]. "Reproduced from R. Agnese et al., Nuclear-Recoil Energy Scale in CDMS II Silicon Dark-Matter Detectors, Nucl. Instruments Methods Phys. Res. Sect. A Accel. Spectrometers, Detect. Assoc. Equip. 905, 71 (2018), with the permission of AIP Publishing."

**Cryogenics**: Assuming that the deposited energy E of a particle in the absorber is fully thermalized, so the temperature rise ΔT is given by:

$$\Delta T = \frac{E}{C_{tot}} \tag{2}$$

Where $C_{tot} = c_V$ is the heat capacity of an absorber with the volume V and the specific heat c. Cryogenic detectors, also known as micro calorimeters, function at low temperatures to exploit the reduced heat capacity of absorber materials, which allows for a significant temperature increase upon particle detection, and are designed with very small absorber volumes for enhanced sensitivity. By applying the Debye model and calculating the internal energy of the lattice vibrations (phonons), the specific heat of a dielectric crystal absorber comes out to be:

$$c_{dielectric} = \beta\left(\frac{T}{\theta_D}\right)^3 \tag{3}$$

With $\beta = 1944\,\mathrm{Jmol^{-1}\,K^{-1}}$ and $\theta_D$ the Debye temperature of the crystal. The cubic dependence on temperature demonstrates a strong decrease of the phonon specific heat at low temperatures [77].

**Background Rejection**: Dominant background events in a detector primarily occurred slightly above the energy threshold of 10 keV [81]. A timing parameter based on phonon pulse rise time and delay compared to the ionization pulse and specific cuts for each detector is used for event rejection [82]. Overall, calibration is performed with gamma and neutron sources, such as $^{133}$Ba and $^{252}$Cf, to tune detectors' responses and set data quality and WIMP detection criteria [82]. Neutron backgrounds, from muons and radioactive processes within the apparatus [83], can be investigated using simulation software, yielding estimated background rates in different energy ranges. Working in any underground facility can reduce the muon flux significantly, and remaining muons can be tagged for rejection by a plastic scintillator muon veto [79]. While cosmic-ray muons are significantly reduced in underground laboratories, other sources must be carefully considered, including:

**Cosmogenic Activation**: As discussed in [84], Germanium and other detector materials undergo cosmogenic activation due to exposure to cosmic-ray neutrons at the Earth's surface. This leads to the production of isotopes such as $^{68}$Ge, $^{57}$Co, and $^{54}$Mn, contributing to background events. Shielding and underground storage before deployment help reduce this effect.

**Radioactive Decay and Environmental Contamination**: Underground experiments still experience low-level radioactivity from uranium and thorium decay chains, potassium-40, and radon gas, which contribute to gamma-ray backgrounds [85].

**Surface and Detector-Induced Events**: Low-energy electron recoils from surface contamination or detector material interactions can be misclassified as WIMP-like events. The CDMS collaboration developed phonon-based discrimination techniques to reject such events with >99.7% efficiency [86].

**Background Mitigation Strategies**: Experiments implement active and passive shielding, including lead, polyethylene, and muon veto systems, along with material selection and purification techniques to minimize intrinsic contamination. Cryogenic and phonon-ionization-based detectors enhance the rejection of electron recoils, as demonstrated by the ZIP and BLIP detector technologies in CDMS [86].

## IV. Experimental Results

In this section, we provide the important results of 4 main experiments by adding our insights into them and providing a perspective on each of them.

### *IV.A. CDMS*

The Cryogenic Dark Matter Search (CDMS) project, initiated in 1996, employs specialized detectors to identify dark matter particles, specifically Weakly Interacting Massive Particles. These detectors, made from germanium and silicon, measure both ionization and phonon signals to distinguish dark matter signals from background noise.

**Early CDMS Experiments:** Utilized a tower setup of ZIP detectors at the Soudan Underground Laboratory (SUL) to detect neutron backgrounds and explore WIMP-nucleon interactions [87]. Notable findings included setting upper limits on WIMP interactions based on their mass and interaction characteristics and the capability to measure neutron backgrounds in both silicon and germanium simultaneously, without any loss in detector backgrounds [88,89].

**CDMS II:** The search for dark matter at SUL, revealed that the minimum spin-independent WIMP-nucleon cross-section is $4\times 10^{-43}\mathrm{cm}^2$ at a WIMP mass of 60 $\mathrm{GeV}/c^2$, and the minimum for the spin-dependent WIMP-neutron cross-section is $2\times 10^{-37}$ $\mathrm{cm}^2$ at 50 $\mathrm{GeV}/c^2$ [66]. The techniques for rejecting surface events have direct implications for improving the signal-to-noise ratio in these experiments. Focused on refining detection sensitivity, particularly for low-mass WIMPs (down to 10 $\mathrm{GeV}/c^2$), calibration with a $^{252}$Cf source indicated no WIMP events, allowing researchers to set stringent upper limits on the spin-independent cross-section for WIMPs (Figure 4). Applying Machine learning algorithms and simulations tools could be levered to distinguish between background noise and potential dark matter signals more effectively. Implementing these tools leading to the establishment of a maximum limit on the interaction cross-section for a WIMP with a mass of 8.6 $\mathrm{GeV}/c^2$, detailed in Figure 4 [79,90–92].

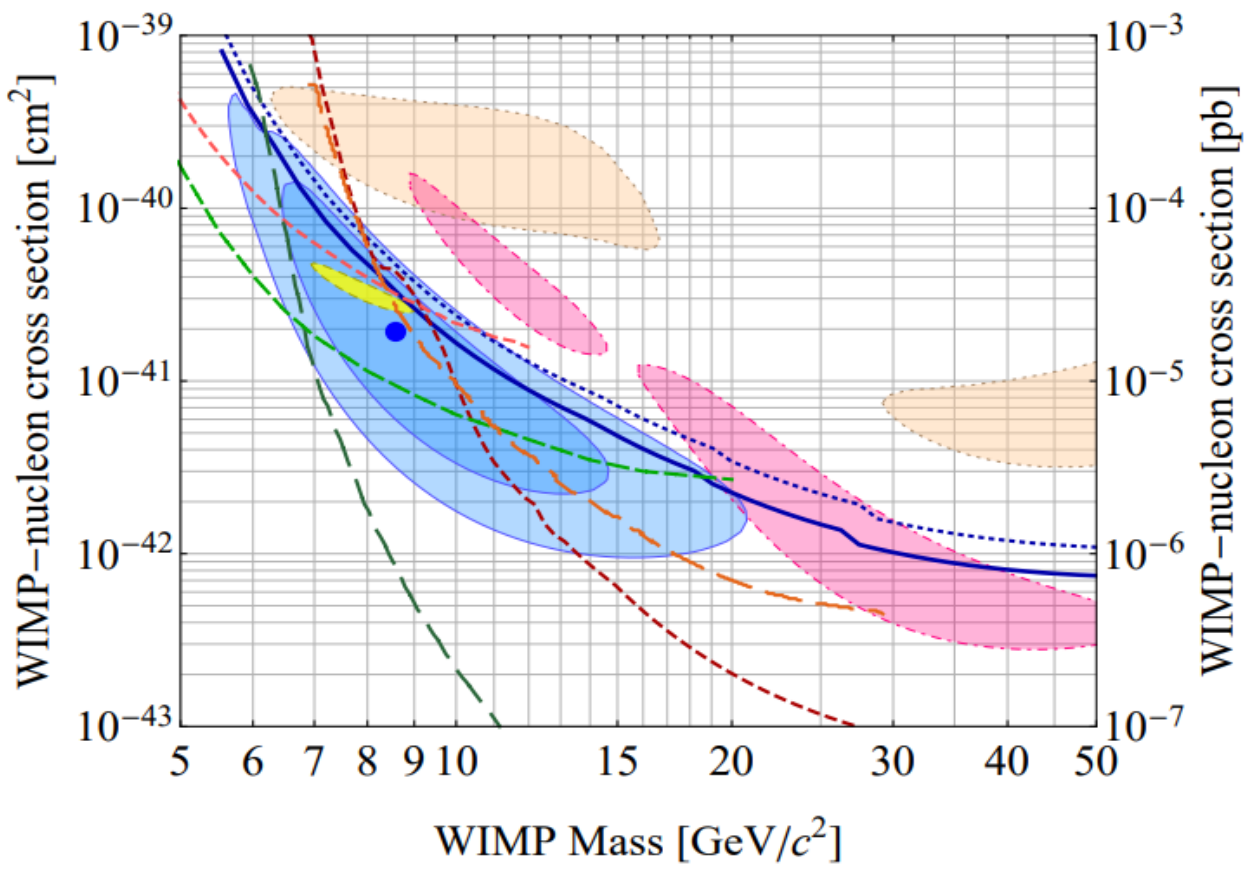

FIG. 4. Experimental data presents upper limits at a 90% confidence level for the spin-independent cross section between WIMPs and nucleons, charted against WIMP mass. The limits are based on the work of Ref. [93,94] (blue dotted line) and when combined with the CDMS II silicon dataset (blue solid line). Additional constraints are shown from CDMS II germanium analyses [95,96] (red dashed lines), EDELWEISS [97] (orange dashed line), and XENON experiments [98,99] (green dash-dotted lines). Areas filled in yellow, tan, and pink indicate potential signals from CoGeNT [100], DAMA/LIBRA [101,102], and CRESST [92,103] experiments, respectively. Light blue contours

represent 68% and 90% confidence levels for a signal from these data alone. A blue dot marks the most likely point for a WIMP signal, with a mass and cross section of 8.6 GeV/$c^2$ and $1.9\times10^{-41}$ cm $^2$, respectively.

**SuperCDMS:** In the pursuit of detecting dark matter particles with masses of 10 GeV/$c^2$or less, SuperCDMS SNOLAB employed two innovative types of cryogenic detectors: HV (high-voltage) and iZIP (interleaved Z-sensitive Ionization and Phonon) detectors. By combining these two technologies, SuperCDMS SNOLAB enhanced its capacity to search for low-mass dark matter candidates and effectively manage expected background noise. The HV detectors are particularly adept at sensing nuclear recoils from dark matter particles with masses below 1 GeV/$c^2$, although they remain sensitive to particles with masses over 5 GeV/$c^2$. Meanwhile, iZIP detectors due to have the advantage of distinguishing between nuclear and electron recoils are designed to provide improved sensitivity for detecting dark matter particles and reduce the major background within above mass range. The ionization signal for iZIP can be normalized to:

$$E_Q = \eta y E_R \tag{4}$$

Where $\eta$ is a position-dependent correction factor that counts for lower measured ionization signals from events near the surfaces and sidewalls of the detector (Table I) and y is ionization yield.

To model the ionization and phonon production of an event in the detector, we consider two classes of events: ER and NR. For the same recoil energy ER deposited in the crystal, ERs generate a larger amount of ionization than NRs. The ratio of the ionization produced by NRs to that of ERs is called the ionization yield. The model treats ionization as a continuous variable, with no statistical fluctuations in the amount of produced ionization. The total phonon signal from an event, $E_{PT}$ is given by:

$$E_{PT} = E_R + E_{Luke} = E_R + \eta \frac{yE_R}{\varepsilon} e\Delta V \tag{5}$$

Where the first term is the recoil energy, and the second term is the additional phonon signal generated through the Luke-Neganov effect ($E_{Luke}$), with $yE_R/\varepsilon$ being the amount of ionization produced, $\varepsilon$ is the average energy required to create an electron-hole pair, and $e\Delta V$ is the work done to move one charge through the crystal. For iZIPs, $E_{Luke}$ is on the order of $E_R$, but in the HV detectors, where $\Delta V$ is large, the Luke term can dominate the phonon signal. Equation 5 treats the ionization as a continuous variable, with no statistical fluctuations in the amount of produced ionization [104].

| Event Location and Type | Ge | Si |
|---|---|---|
| Bulk Events | 1.0 | 1.0 |
| Events near the top/bottom faces | 1.0 | 1.0 |
| Events near the cylindrical sidewalls | 0.75 | 0.90 |
| ERs on the top/bottom faces | 0.70 | 0.65 |
| ERs on the cylindrical sidewalls | 0.52 | 0.585 |
| $^{206}$Pb recoils on the top/bottom faces | 0.65 | 0.65 |
| $^{206}$Pb recoils on the cylindrical sidewalls | 0.488 | 0.585 |

TAB I. Fraction of ionization collected by events of given locations and types. "Reproduced with permission from R. Agnese et al., Projected Sensitivity of the SuperCDMS SNOLAB Experiment, Phys. Rev. D 95, 082002 (2017). Copyright 1998 American Physical Society."

*IV.A.1. Main features of CDMS detectors*

Here, the main features of these detectors which make them prominent, are presented.

**Ionization Yield Models:** Utilizing the Lindhard theory and its modifications to better predict ionization yields for nuclear recoils is a critical factor in identifying valid dark matter events. The energy loss of charged particles traveling through a material consists of two main components: electronic stopping and nuclear stopping, each exhibiting distinct energy dependencies. Slow-moving nuclear recoils interact primarily with the target's atomic nuclei, as they are not effectively slowed down by electronic interactions. Consequently, most of their energy is deposited through nuclear collisions rather than ionization. Since ionization results from electronic excitations, nuclear recoils produce a significantly lower ionization yield than electron recoils of the same energy [105]. The ionization yield of nuclear recoils is influenced by how energy is partitioned between electronic and nuclear interactions in the stopping process. The expected ionization yield for a nuclear recoil under Lindhard theory is given by:

$$y_L = \frac{kg(\varepsilon)}{1 + kg(\varepsilon)} \tag{6}$$

By defining Z as the atomic number and A as the atomic mass, the parameter is $k = 0.133Z^{2/3} A^{-1/2} \approx 0.146$ for silicon, and the transformed energy $\varepsilon = 11.5E_R\, Z^{-7/3}$, with the recoil energy $E_R$ given in keV. The function $g(\varepsilon)$ is well-fit by a polynomial in $\varepsilon$ with empirically chosen coefficients, described by $3\varepsilon^{0.15} + 0.7\varepsilon^{0.6} + \varepsilon$ [106]. Standard Lindhard theory significantly over-estimates the ionization production for low energy nuclear recoils so, an improved functional form using a parameter a = 0.247 matches the Lindhard expectation $y_L$ for silicon at high energy and fits the data reported in Ref. [107] at low energy (Figure 5):

$$y_C = \left(\frac{1}{aE_R} + \frac{1}{y_L}\right)^{-1} \tag{7}$$

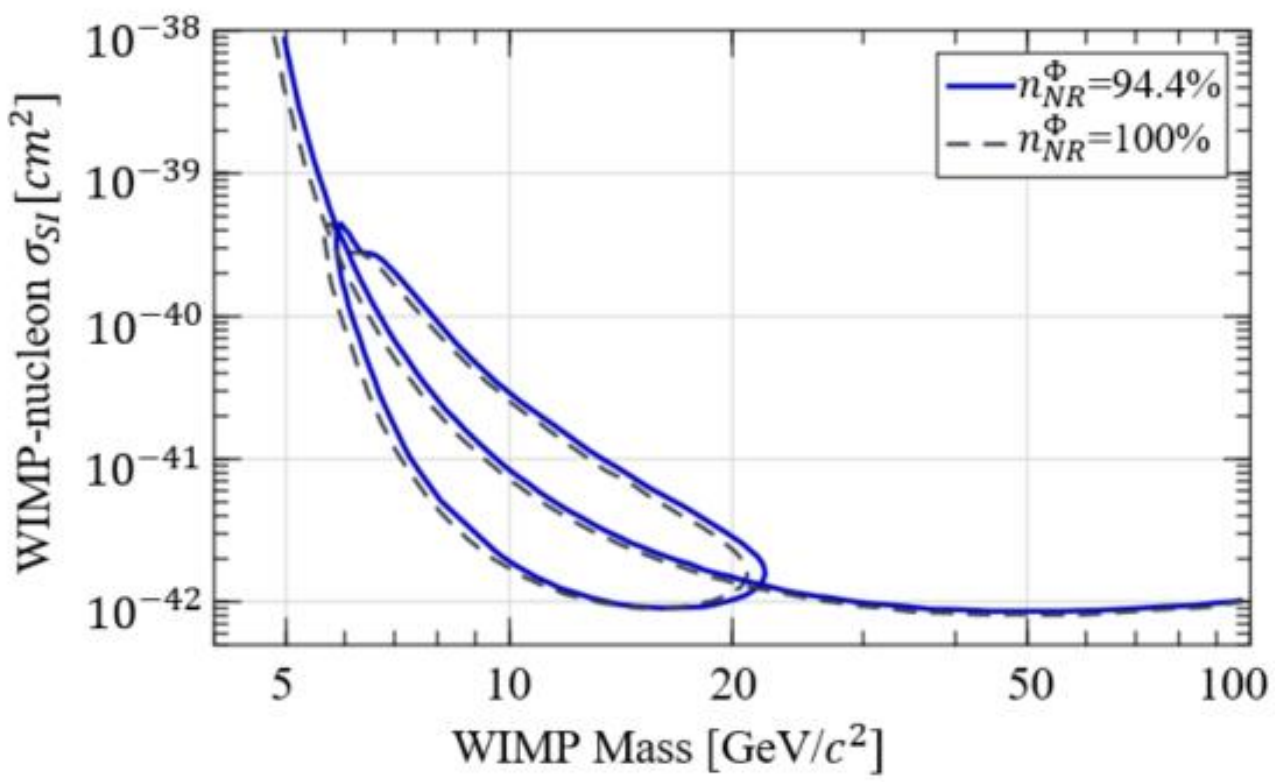

FIG. 5. Updates to the nuclear-recoil energy scale only marginally affect the WIMP detection sensitivity and contours from the CDMS II data in Ref. [108]. As shown here, there are minor shifts in the spin-independent WIMP-nucleon cross-section exclusion curve and the region of best-fit WIMP mass and cross-section, both at a 90% confidence level. The shift is less than 20% for WIMPs heavier than 10 GeV/$c^2$, while the upper limit for those around 5 GeV/$c^2$ roughly doubles. Additionally, the best-fit WIMP mass shifts by less than 5% due to the revised energy scale [80]. "Reproduced from R. Agnese et al., Nuclear-Recoil Energy Scale in CDMS II Silicon Dark-Matter Detectors, Nucl. Instruments Methods Phys. Res. Sect. A Accel. Spectrometers, Detect. Assoc. Equip. 905, 71 (2018) with the permission of AIP Publishing."

**Detector Efficiency:** Noted that ionization yield efficiency decreases with increasing recoil energy as the self-shielding effect, affecting the detector's ability to accurately record higher energy events which the efficiency of detector diminishes for higher energy, dropping to approximately 75%.

**Explore New Dark Matter Candidates:** Beyond traditional WIMPs, experiments are beginning to consider other potential dark matter candidates like dark photons, which could interact through different mechanisms. For DM particles as light as 1 MeV/$c^2$, detection methods such as electron scattering and dark photon absorption become relevant, and these are being probed with a prototype SuperCDMS detector that has a high charge resolution. It is expected large electric fields in this context can cause unwanted ionization and charge tunneling, which necessitates monitoring the charge leakage rate to filter out noise. The theoretical framework for detecting dark photons involves assuming, they kinetically mix with standard model photons. The subsequent interaction of the Standard Model (SM) photon with the material was computed according to tabulated photoelectric cross sections [109], giving the approximate event rate [110]:

$$R = V_{det} \frac{\rho_{DM}}{m_V} \varepsilon_{eff}^2 (m_V, \tilde{\sigma}) \sigma_1 (m_V) \tag{8}$$

Where $V_{det}$ is the detector volume, $\rho_{DM}/m_V$ is the number density of DM which is ~ 0.3 $GeVc^{-2}cm^{-3}$, $m_V$ is the dark photon mass, $\varepsilon_{eff}$ is the effective kinetic mixing angle [110], $\tilde{\sigma}$ is the complex conductivity, and $\sigma_1(m_V)$ = $Re(\tilde{\sigma}(m_V))$ is computed from the photoelectric cross section $\sigma_{p.e.}$ [109]. Using an ionization production model to project an absorption event of known energy into measured signal space [111–113], the mean $n_{eh}$ can be calculated:

$$\langle n_{eh}(E_\gamma)\rangle = \begin{cases} 0 & E_\gamma < E_{gap} \\ 1 & E_{gap} < E_\gamma < \epsilon_{eh} \\ E_\gamma/\epsilon_{eh} & \epsilon_{eh} < E_\gamma \end{cases} \tag{9}$$

Where $E_{gap} = 1.12$ eV and $\epsilon_{eh} = 3.8$ eV [114]. The signal induced by ERDM (Electron Recoil Dark Matter) was calculated according to the formalism in Ref. [115] in which scattering rates accounting for band structure in Si are tabulated for signal modeling. The differential scattering rate is given by the function:

$$\frac{dR}{d\,lnE_R} = V_{det}\frac{\rho_{DM}}{m_{DM}}\frac{\rho_{Si}}{2m_{Si}}\sigma_e\alpha\frac{m_e^2}{\mu_{DM}^2}I_{crystal}(E_e;F_{DM}) \tag{10}$$

Where $\sigma_e\alpha$ encodes the effective DM-SM coupling, $F_{DM}$ is the momentum transfer (q) dependent DM form factor, $\mu_{DM}$ is the reduced mass of the DM-electron system, and $I_{crystal}$ is the scattering integral over phase space in the crystal [115]. To obtain the anticipated quantized spectrum, the differential spectrum (Equation 10) is integrated with Equation 9, with the application of the same energy resolution smearing used for dark photon signals. The resulting quantized signals, generated by Electron Recoil Dark Matter (ERDM) and dark photons, are selected through the optimum interval method to match the data that resemble these signals [109].

**Inelastic Scattering:** In the elastic scattering of DM, dark matter particles transfer mostly of its momentum to the target with a similar mass, then the energy of the recoiling nucleus is:

$$E_r = \frac{|q|^2}{2m_N} = \frac{m_\chi^2 v^2}{2m_N} \tag{11}$$

Where $m_N$ is the mass of the nucleus, $m_\chi$ is the dark matter mass, $v$ is the velocity of the dark matter, and $q = m_\chi v$ is the momentum of the incoming dark matter particle. Equation 11 indicates that the energy of recoil is directly related to the square of the mass of the dark matter particles; hence, detecting the elastic recoil energy of low-mass dark matter particles is exceedingly challenging. If we consider that a scattering process could involve the creation of an additional particle, carrying away the maximum possible energy, which would be the total energy of the system, then this scenario can be expressed as follows:

$$E_\chi = \frac{1}{2}m_\chi v^2 \tag{12}$$

Combining Equation 11 and 12 gives:

$$E_r = E_\chi\frac{m_\chi}{m_N} \tag{13}$$

Assuming we are focusing on the low-mass dark matter regime where the mass of the dark matter particle $m_\chi$ is much smaller than the mass of the nucleus $m_N$, the recoil energy $E_R$ will be significantly less than the dark matter particle's kinetic energy $E_\chi$. This means that for a low-mass dark matter particle, an inelastic scattering event will produce a larger detectable energy signature compared to an elastic scattering event. By increasing the recoil energy detection threshold to 150 keV and enhancing the methods for rejecting surface event backgrounds, three potential dark matter events were identified in the energy range of 25 keV to 150 keV in CDMSlite Round 3 (as illustrated in Figure 6). The likelihood of observing three or more events attributable to background noise using the background distribution $f(\mu)$ is calculated 11% using Equation 14 which, while relatively low, is still significant enough that it cannot be disregarded [83].

$$p = \int_0^\infty d\mu \, f(\mu) \cdot \sum_{k=3}^{\infty} \frac{e^{-\mu}\mu^k}{k!} \tag{14}$$

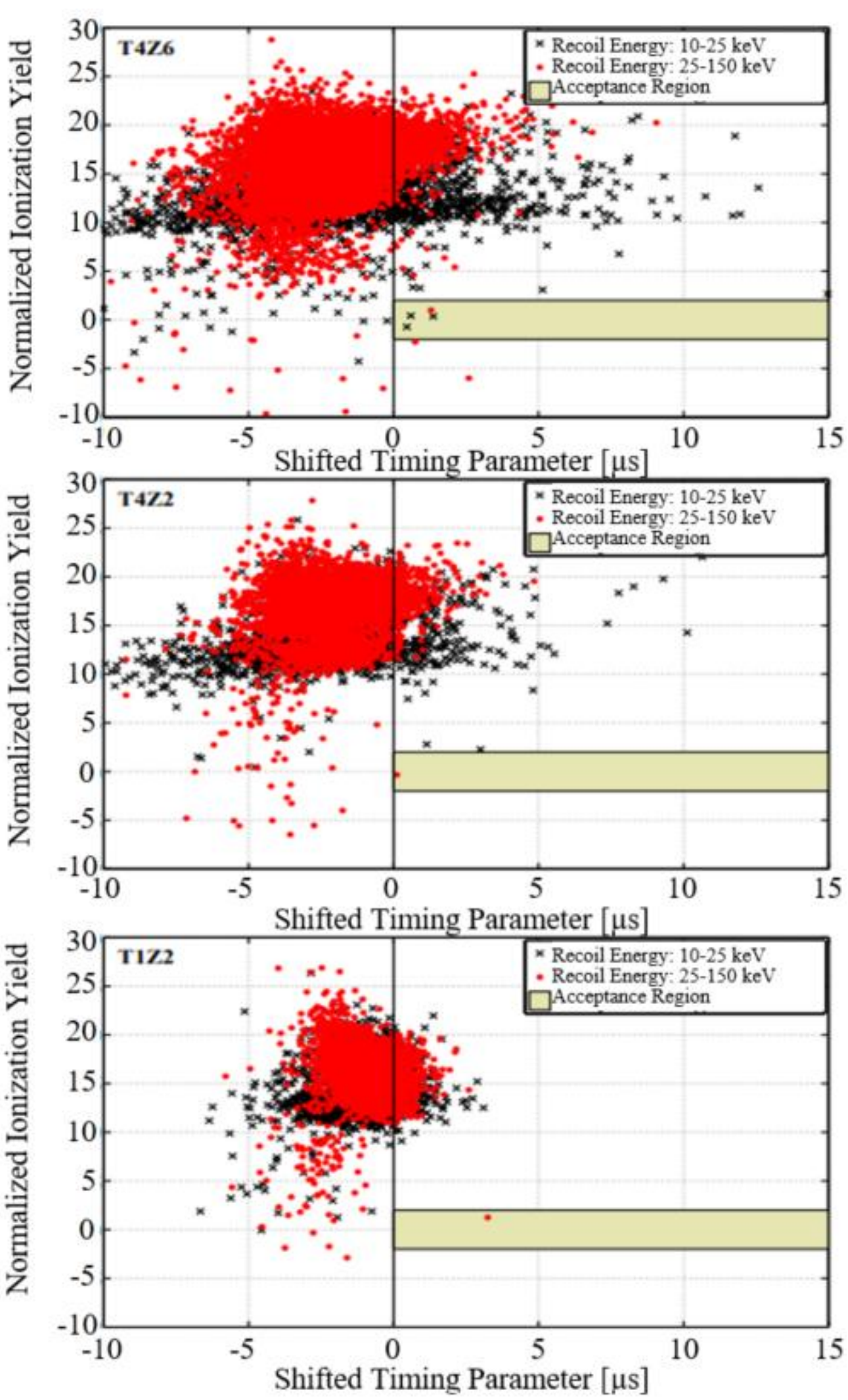

FIG. 6. The data presents how far each event deviates from the average nuclear-recoil signal, in terms of standard deviations, when considering the ionization yield and adjusted timing parameters, for three detectors that identified potential Weakly Interacting Massive Particles (WIMPs) with energies exceeding 25 keV. Specifically, the detectors T4Z6, T4Z2, and T1Z2 reported WIMP candidate events at energy levels of 37.3 keV, 73.3 keV, and 129.5 keV, respectively. Additionally, the T4Z6 detector detected three more potential WIMP events within the lower energy bracket of 10–25 keV. The regions of the graph that are considered valid for WIMP candidate events are marked by shaded areas [83]. [80]. "Reproduced from Z. Ahmed et al., Search for Inelastic Dark Matter with the CDMS II Experiment, Phys. Rev. D - Part. Fields, Gravit. Cosmol. 83, 112002 (2011) with the permission of AIP Publishing."

### *IV.A.2. Insights on CDMS Collaboration*

CDMS collaboration due to the use of multiple types of detectors, including ZIP, BLIP, iZIP, and HV detectors, shows a robust approach to capturing various interactions and enhancing the sensitivity to dark matter particles. The ability to measure neutron backgrounds simultaneously in both silicon and germanium without loss of detector performance is a considerable achievement, helping to reduce background noise which is crucial for searching the rare events. The project has successfully set more stringent upper limits on the spin-dependent and spin-independent cross-sections of WIMP-nucleon interactions, pushing forward the boundaries of dark matter particle detection. Extensive calibration and simulations have been effectively used to refine the detection capabilities and filter out potential false signals.

However, background noise and false signals**,** despite advances is still the issue which poses a significant challenge, as seen with the statistical likelihood of misinterpreting background noise as potential WIMP detections. The absolute energy scale necessary for nuclear recoils and the various corrections required for accurate data interpretation add layers of complexity and potential error to the analyses. The decrease in ionization collection efficiency for nuclear recoils over 20 keV limits the detector's effectiveness in higher energy ranges.

We expect enhanced detector materials and design could provide better efficiency at higher energies and improved sensitivity at lower masses. Also, developing more sophisticated calibration techniques that can more accurately mimic the expected dark matter signals could help in further refining the experimental setups. Continued efforts to minimize and understand background noise through better shielding, isolation, and background modeling are crucial.

## ***IV.B. EDELWEISS***

EDELWEISS (**E**xpérience pour **DE**tecter **L**es **W**IMPs **E**n **S**ite **S**outerrain) is a dark matter search experiment located at the Modane Underground Laboratory in France. The experiment employs cryogenic detectors that simultaneously measure the phonon and ionization signals resulting from particle interactions within germanium crystals. These detectors are made from high-purity germanium mono-crystals and function as bolometers. They are fitted with Neutron Transmutation Doped (NTD) Germanium sensors that act as thermistors to detect temperature changes. Additionally, the detectors are equipped with concentric aluminum ring electrodes [116].

The initial EDELWEISS experiments started with an exposure of 62 kg-days and reduced the detector threshold from 30 keV to 13 keV by employing a phonon trigger [117–119]. Implementing Monte-Carlo simulations using Geant4 to analyze gamma and neutron backgrounds from radioactive decays and upgrading both electronics (readout and DAQ) and cryogenic systems, along with improved shielding and the use of newly designed Fully Inter-Digitized (FID) detectors achieved a rejection power of $R_{surf} < 4 \times 10^{-5}$ per α at 90% confidence level using a $^{210}Pb$ source. Furthermore, the experiment was capable of rejecting bulk γ-ray events using γ-calibrations with a $^{133}Ba$ source, resulting in a misidentification rate $R_{\gamma-mis-FID} < 2.5\times10^{-6}$ at 90% confidence level [120–124].

### *IV.B.1. Insights on EDELWEISS*

Utilizing the Migdal effect and NTL effect to target dark matter particles as light as 32 MeV/c² indicates a strategic shift towards exploring a broader range of dark matter masses. Achieving a record sensitivity of 0.53 electron-hole pairs sets a new standard in detector resolution. Continuous improvements in this area could enable the detection of even subtler signals from dark matter-electron interactions, enhancing the likelihood of detecting dark photon dark matter down to 1 eV/$c^2$ and other elusive candidates (Figure 7).

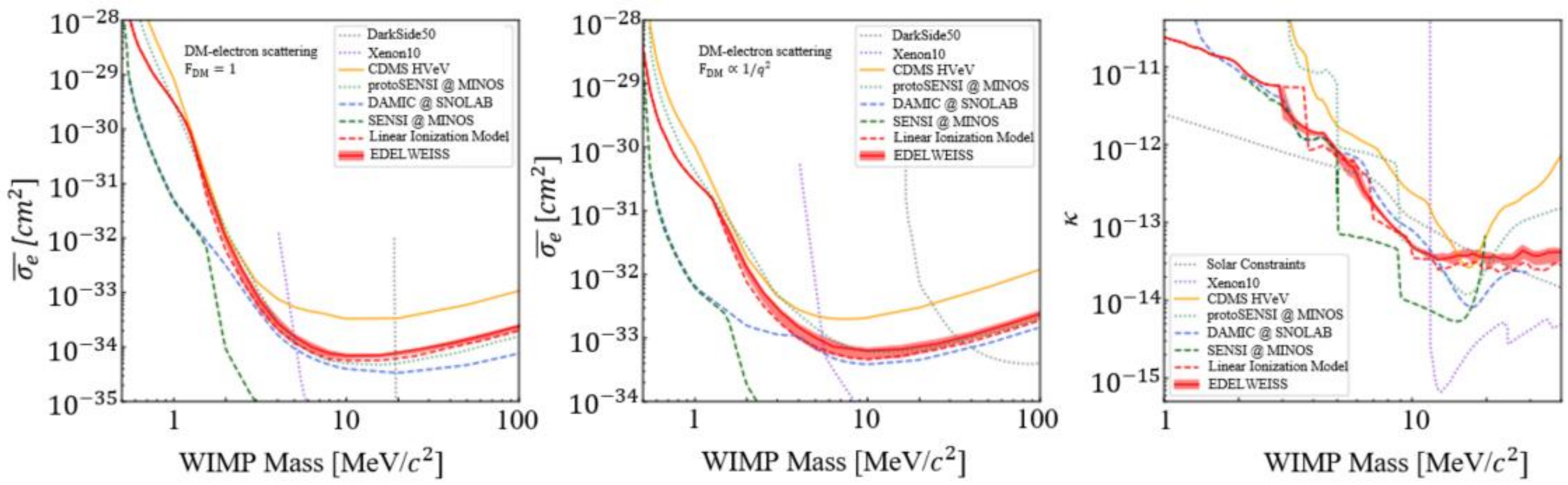

FIG. 7. The 90% confidence level (C.L.) upper boundary for the interaction cross-section between dark matter (DM) particles and electrons is presented under two different assumptions: a heavy mediator (shown in the left panel) and a light mediator (depicted in the middle panel). The right panel displays the upper boundary on the kinetic mixing parameter κ associated with a Dark Photon [125]. The shaded red band and the dotted red line depict varying models of charge distribution as detailed in Reference [126]. Also constraints from other direct detection experiments are shown [109,127–131].

### ***IV.C. CoGeNT***

The CoGeNT (Coherent Germanium Neutrino Technology) collaboration uses p-type point-contact (PPC) germanium detectors in their search for WIMPs. They reported measurements of a spectrum consistent with nuclear recoil events and a temporal variation in these events that aligns with predictions for a light dark matter particle with a mass between 4.5 and 12 GeV/$c^2$ elastically scattering with nuclei. The implied cross-section with nucleons is approximately $\sim 10^{-40}$ cm$^2$.

CoGeNT's findings are also in agreement with the annual modulation signature reported by the DAMA/LIBRA collaboration [132]. Significant advancements in shielding, data acquisition, instrumental stability, data analysis, and background estimation contributed to a pivotal result that is a detailed assessment of surface event contamination in the dataset as a function of energy, which is illustrated in Figure 8.

*IV.C.1. Insights on CoGeNT*

Notable characteristics of these detectors include: (a) the relative straightforwardness of CoGeNT's data analysis, leading to consistent irreducible spectra across different analysis methods; (b) a well-understood response to nuclear recoils, ensuring a reliable scale for nuclear recoil energy; and (c) the capacity for PPC detectors to operate stably without interruption over extended periods, which could span several years, making them particularly apt for detecting the annual modulation that would be indicative of dark matter particles constituting a galactic halo.

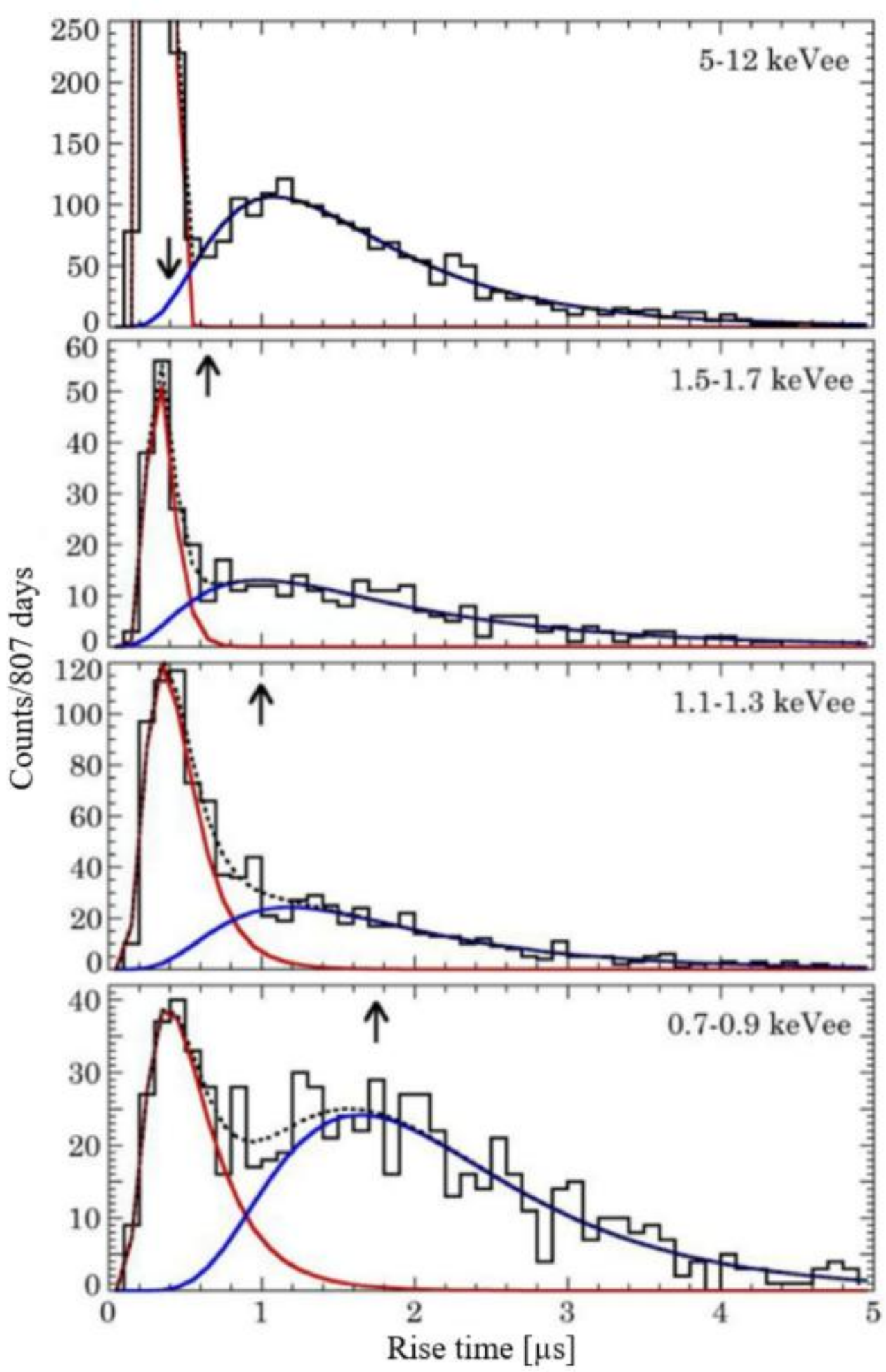

FIG. 8. Rise time distributions for events within specific energy ranges, based on a 27-month data collection period from the CoGeNT detector operating at the Soudan Underground Laboratory (SUL). These distributions are modeled using two separate log-normal distributions, each with adjustable parameters. These two distributions distinguish between surface events with slower rise times (indicated in blue) and bulk events with faster rise times (shown in red). Small vertical arrows mark the threshold for 90% confidence level (C.L.) acceptance of the fast signal, as established by calibration with electronic pulsers. It is noted that as the energy of the events decreases, there is an increasing contamination by surface events that are not effectively rejected, even after applying the fast signal acceptance cut [100]. “Reproduced from C. E. Aalseth et al., CoGeNT: A Search for Low-Mass Dark Matter Using p-Type Point Contact Germanium Detectors, Phys. Rev. D - Part. Fields, Gravit. Cosmol. 88, (2012). Copyright 1998 American Physical Society.”

### *IV.D. DAMIC*

The DAMIC experiment uses Charge-Coupled Devices (CCDs) to detect dark matter particles under 10 GeV, leveraging their low electronic readout noise for precise ionization signal measurements. Operating CCDs at temperatures below 175 K, DAMIC achieves a detection threshold of less than 0.5 keV in nuclear recoil energy, ideal for identifying low-mass dark matter [133]. The experiment's CCDs offer high spatial resolution, essential for analyzing the size and depth of particle interactions within silicon, which aids in distinguishing background noise from valid signals [134,135]. DAMIC initially reported 54 events below 7 $\mathrm{keV_{ee}}$, aligning with expected backgrounds and setting exclusion limits for dark matter interactions (referred to as Figure 9). The next phase, DAMIC-M, enhances detection by using skipper amplifiers for more accurate charge measurements in CCD pixels and plans to deploy a kilogram-scale CCD array in an ultra-low background environment in France, targeting dark matter masses below 1 GeV/$c^2$ with improved sensitivity [136,137].

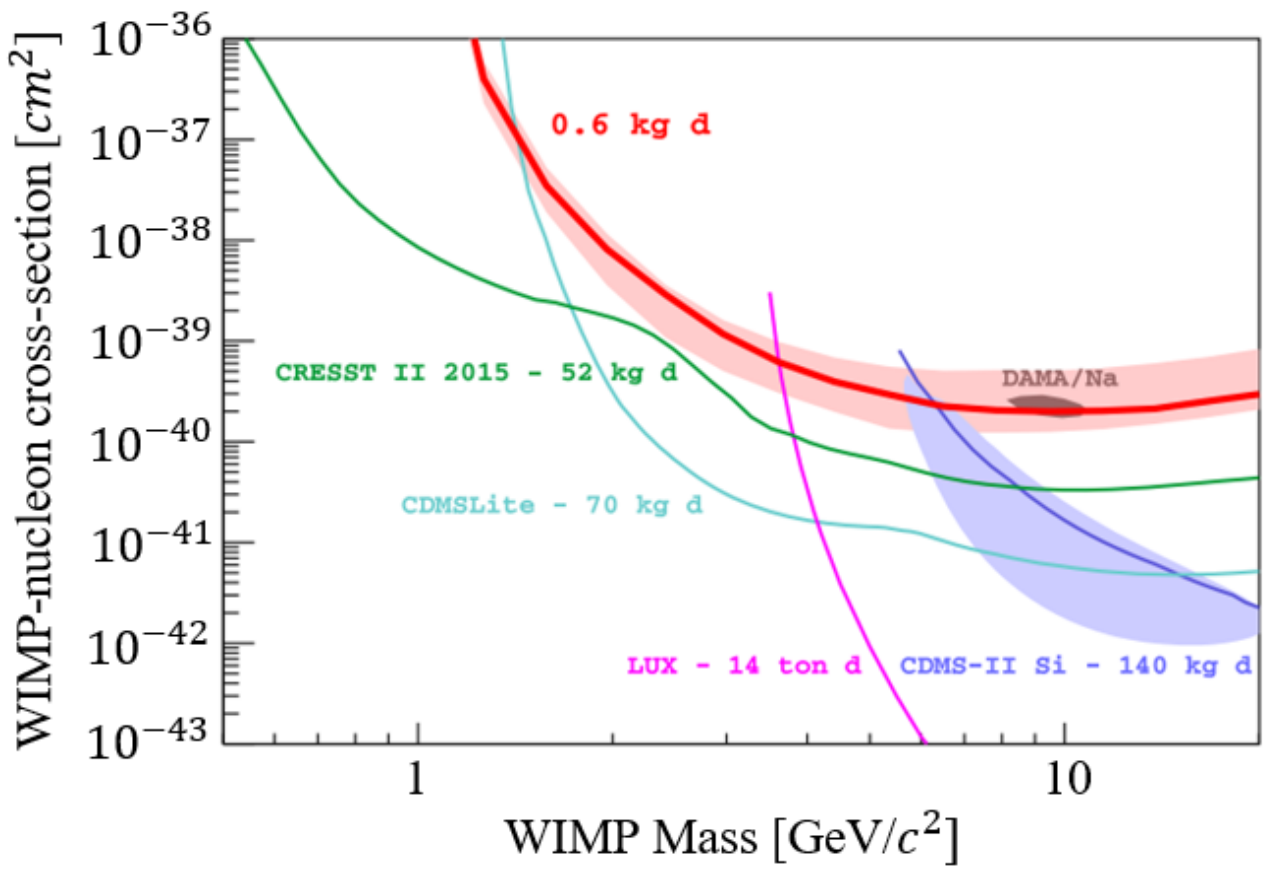

FIG. 9. 90% confidence level (C.L.) upper limit on the WIMP-nucleon cross section is depicted by a red line (0.6 kg d exposure of the DAMIC) [134]. The red band represents the anticipated 1σ sensitivity range. Additionally, for context, the exclusion limits from other experiments at the 90% C.L. [138,139], are provided alongside the 90% C.L. contours that are associated with potential WIMP detection signals from the CDMS-II Si experiment [108] and the DAMA experiment [90]. “Reproduced with permission from A. Aguilar-Arevalo et al., Search for Low-Mass WIMPs

in a 0.6 Kg Day Exposure of the DAMIC Experiment at SNOLAB, Phys. Rev. D 94, (2016). Copyright 1998, American Physical Society."

### *IV.D.1. Insights on DAMIC*

This experiment enables extremely sensitive measurements of ionization signals, with detection thresholds as low as 0.5 keV in nuclear recoil energy. Employing CCDs allows DAMIC to benefit from superior spatial resolution, which helps in accurately determining the size and location of pixel clusters from particle interactions. The ability to distinguish and remove background noise, especially from events occurring near the surface of the device, significantly enhances the reliability of the data collected, and leads to more accurate results. Using skipper amplifiers and plans for a kilogram-scale CCD array, shows promise for further improvements in sensitivity and noise reduction, making the experiment more scalable and capable of exploring new physics.

We believe exploring more efficient cooling systems that can maintain lower temperatures without significant increases in operational costs could improve the performance of CCDs and reduce thermal noise further.

In Table II, we have summarized the last results of the main important experiments which are presented in this section.

| Experiment | Upper limit of DM-Nucleon Cross Section $(\mathbf{cm^2})$ | Lowest Expected DM Mass to Detect $(\mathbf{GeV/c^2})$ | Special Feature and Result |
|---|---|---|---|
| CDMS | $\sim 1.9 \times 10^{-41}$ | 8.6 | 1) Using Simulation tools 2) Three candidates' events which by 81% can be attributed to WIMP signal 3) Detection threshold ~ 1 keV |
| SuperCDMS | $\sim 10^{-28}$ | 93 | 1) Using iZIP and HV detector simultaneously 2) Ability to search for DM particles as light as 1 MeV/$c^2$ |
| EDELWEISS | $\sim 10^{-29}$ | 0.032 | 1) Ability to search for Dark Photon Dark Matter down to 1 eV/$c^2$ 2) Reject surface β- and α-decays |
| CoGeNT | $\sim 10^{-40}$ | 4.5 | Operate stably without interruption over extended periods |
| DAMIC | $\sim 5 \times 10^{-36}$ | 1 | 1) Low electronic readout noise 2) Operates at low temperatures ~175 K 3) Detection threshold ~ 0.5 keV |

TAB II. The last achievements of different experiments which use Ge and Si detectors

## V. Future Prospects

### *V.A. Next-Generation Detectors*

**SuperCDMS:** Earlier sections highlighted key findings from prior iterations of this experiment. The experiment's primary objective is to reach the sensitivity level of the solar neutrino floor. This will be made possible through research and development efforts focused on creating improved detectors that have both lower energy thresholds and reduced background levels [140]. Exploring the use of various target materials such as **L**iquid **He**lium (LHe), **Ga**llium **Ar**senide (GaAs), or Sapphire ($AL_2O_3$), all equipped with phonon sensors, represents a novel research direction outlined in reference [141]. In addition, efforts to identify and mitigate low-energy background events in the HVeV detector have led to the design of a new detector holder. This holder aims to minimize the presence of insulator materials within the detector's volume, thereby reducing luminescence caused by the holder materials themselves, as described in reference [142].

**EURECA:** European Underground Rare Event Calorimeter Array (EURECA) project is an upcoming dark matter search that builds on the work of the CRESST, EDELWEISS, and ROSEBUD experiments. It aims to detect scalar interaction cross sections in the range of $10^{-45}$ – $10^{-46}$ cm$^2$ and plans to use up to one tonne of target mass [143]. EURECA's multi-target approach is advantageous for confirming the expected atomic mass scaling of WIMP-nucleon interactions and identifying any neutron background. The diversity of target materials also helps to address the uncertainty in WIMP mass. The project's main challenge is to enhance the detectors in terms of size, sensitivity, and purity, as well as to tackle the logistics of mass production [144].

**Silicon Photomultipliers (SiPMs):** The Silicon Photomultiplier (SiPM) is a solid-state sensor capable of detecting low-light levels, like the traditional Photomultiplier Tube (PMT). SiPMs offer several advantages over PMTs, including operation at low voltages, immunity to magnetic fields, mechanical durability, and consistent response across the sensor [145]. SiPMs are considered promising for use in future dark matter experiments involving liquid xenon (LXe) due to their low radioactivity [146]. However, a significant challenge for their broader adoption is their high Dark Count Rate (DCR), which needs to be reduced to match the lower levels found in PMTs (approximately 0.01 Hz/mm$^2$) currently used in LXe experiments [147].

### *V.B. Collaborative Efforts*

Recent progress in collaborative dark matter searches has greatly enhanced the constraints on the WIMP-nucleon cross-section. By merging their datasets, the EDELWEISS and CDMS collaborations achieved a total exposure of 614 kg·day, resulting in a more restrictive upper limit of $3.3 \times 10^{-44}$ cm$^2$ for a 90 GeV/c$^2$ WIMP in the spin-independent interaction channel. The Optimum Interval method was applied to improve background rejection, yielding stronger constraints at lower WIMP masses while setting tighter limits at higher masses, leveraging differences in the energy spectra of both experiments [123]. This combined analysis enhances the statistical power of dark matter detection and helps refine exclusion limits for various WIMP mass ranges.

Moreover, the EURECA project, a joint effort involving EDELWEISS and CRESST, aims to push these limits further by employing a multi-target approach, utilizing both germanium and $CaWO_4$ detectors [143]. EDELWEISS employs charge-phonon detectors to measure ionization and phonon signals simultaneously while CREST utilize scintillation-phonon detectors to measure both phonons and scintillation. Combined analysis increased the probability to distinguish signal events from background and allowing better identification of nuclear recoil in the presence of electron-recoil backgrounds. By scaling the target mass up to one ton, EURECA intends to explore WIMP cross-sections in the $10^{-45}$ to $10^{-46}$ $\mathrm{cm}^2$ range, providing an unprecedented level of sensitivity. This multi-material strategy enables cross-verification of nuclear recoil signals and minimizes background contamination.

These collaborative efforts highlight the necessity of combining data from multiple experiments to achieve higher sensitivity and improve dark matter search methodologies, significantly advancing our understanding of WIMP interactions.

### *V.C. Theoretical and Computational Advances*

#### *V.C.1. Simulation Software*

GEANT4 and FLUKA are both software packages used for simulating the passage of particles through matter, often utilized in particle physics, nuclear experiments, and astrophysics for detector design, and understanding background and signal in experiments. GEANT4 supports HepMC, a C++ Monte Carlo event record, for handling primary particles in simulations, variance reduction techniques such as importance biasing, weight-window methods, and cross-section biasing are implemented, particularly for radiation shielding studies, predefined Physics Lists, allowing users to select suitable Monte Carlo-based physics models, including Binary Cascade, Bertini Cascade, CHIPS, and Penelope for hadronic and electromagnetic interactions. FLUGG (FLUKA with GEANT4 Geometry) allows FLUKA to use GEANT4's geometry package, enabling simulations with GEANT4-based geometry inputs [148–151]. These simulation tools might be applied to:

**Model the CDMS II detector:** Simulate the detector response to particle interactions, including nuclear recoils that are expected from dark matter interactions. This can help in designing and optimizing the detector setup to maximize sensitivity to dark matter signals while minimizing the backgrounds.

**Understand background:** Simulate the various components of the background, such as cosmic rays, natural radioactivity, or other sources that could mimic the expected signal from dark matter particles.

**Optimize signal discrimination**: Help in developing and testing methods for distinguishing between background (such as electron recoils) and genuine dark matter signals (like nuclear recoils).

**Predict experimental outcomes:** Provide predictions of the expected number of events from dark matter interactions, which can then be compared with the actual data to look for excesses that may indicate a discovery or to set limits on dark matter properties.

*V.C.2. Data Mining*

Machine learning has emerged as a transformative tool in dark matter research, addressing the challenge of extreme class imbalance in the search for elusive particles such as dark photons and neutralinos [152]. It has proven to enhance detection efficiency and significantly reduce computational demands by 30 times compared to conventional methods, without sacrificing accuracy [153]. Machine learning algorithms adeptly manage non-linear variable correlations and incorporate additional data types, like light signal pulse shapes, into analyses. This leap in efficiency and capability allows for more sophisticated experimental models and easier integration of complicating factors into analyses, as evidenced by its application in the SHiP experiment to discriminate dark matter signals from prevalent neutrino backgrounds [154].

Overall, machine learning stands out as a sophisticated instrument in the quest for dark matter and supersymmetry, demonstrating its capability to manage intricate data and to pinpoint signals that traditional techniques typically struggle to discern.

### *V.D. Challenges and Limitations*

The ability to detect exceedingly uncommon events indicative of dark matter interacting with the detector's substance is dependent on the detector's high level of sensitivity. As previously mentioned, Ge and Si detectors are plagued by various background interferences, such as cosmic rays and terrestrial radioactivity. To augment the probability of identifying authentic dark matter interactions, these detectors are often fortified with advanced shielding, situated in subterranean facilities, and analyzed using intricate data processing techniques. The financial outlay for Ge and Si detectors is considerable, covering a range of costs from acquiring and refining the materials to constructing the detectors and the necessary infrastructure for their operation, which includes deep underground settings. Additionally, the investment in computational and manpower for data scrutiny is significant. The operational efficacy of these detectors in an environment with minimal background interference necessitates substantial investment in cutting-edge technology and rigorous standards of cleanliness.

## VI. Conclusion

In this comprehensive review, we have explored the pivotal role of Germanium and Silicon detectors in the ongoing quest to detect and understand dark matter, particularly focusing on Weakly Interacting Massive Particles (WIMPs). These detectors, renowned for their minimal radioactive backgrounds and substantial active volumes, have significantly advanced the frontier of dark matter research. Through the lens of various experiments and collaborations, we have seen how the unique properties of these detectors enable the precise measurement of nuclear recoils, offering a promising avenue towards the direct detection of dark matter particles.

Moreover, the advancements in detector technology, particularly the development of cryogenic detectors and their enhanced sensitivity at low temperatures, represent a monumental step in measuring interactions that were previously beyond our reach. The integration of sophisticated electronics and phenomenological models has enabled these detectors not only to capture rare particle interactions but also to discriminate between different types of particle interactions, thereby reducing background noise and enhancing signal clarity.

As we look forward, the field of dark matter detection stands on the brink of potentially transformative discoveries. The ongoing refinement of detector technologies and experimental techniques promises to increase our sensitivity to possible dark matter signals. Furthermore, the theoretical exploration of dark matter candidates continues to evolve, suggesting new pathways for investigation. The collaborative efforts across international borders underscore the global importance of this search, highlighting the universal pursuit to unravel the mysteries of the cosmos. The journey to detect and understand dark matter is fraught with challenges yet ripe with opportunity. The continued use of Germanium and Silicon detectors, coupled with theoretical insights and technological advancements, holds the potential to eventually uncover the nature of dark matter. Such a discovery would not only answer long-standing questions about the composition of the universe but also redefine our understanding of its fundamental physical laws.